\documentclass[a4paper,10pt]{article}
\usepackage[utf8]{inputenc}
\usepackage{amsmath}
\usepackage{hyperref}
\usepackage{graphicx}
\usepackage{amsmath,amssymb,amsthm} 
\usepackage{enumerate}
\usepackage{amsthm}

\theoremstyle{plain} 
\newtheorem{theorem}{Theorem}[section]

\newtheorem{proposition}[theorem]{Proposition}
\newtheorem{corollary}[theorem]{Corollary}

\theoremstyle{definition}
\newtheorem{definition}{Definition}[subsection]

\theoremstyle{remark}
\newtheorem{remark}{Remark}

\DeclareGraphicsExtensions{.pdf,.png,.jpg}
\title{On a capital allocation principle coherent with the solvency 2 standard formula.}
\author{Fabio Baione\\
University of Florence\\
Department of Economics and Management\\
Via delle Pandette 9, 50127 Florence, Italy\\
Email: \texttt{fabio.baione@unifi.it}
\medskip\\
Paolo De Angelis\\ 
Sapienza University of Rome\\ 
Via Del Castro Laurenziano 9, 00161 Rome, Italy\\
Email: \texttt{paolo.deangelis@uniroma1.it}
\medskip\\
Ivan Granito\\
Sapienza University of Rome\\
Viale Regina Elena 295/G, 00161 Rome, Italy\\
Email: \texttt{ivan.granito@uniroma1.it}
}
\begin{document}
\maketitle

\begin{abstract}
Solvency II Directive 2009/138/EC requires an insurance and reinsurance undertakings assessment of a Solvency Capital Requirement by means of the so-called “Standard Formula” or by means of partial or full internal models. Focusing on the first approach, the bottom-up aggregation formula proposed by the regulator permits a capital reduction due to diversification effect, according to the typical subadditivity property of risk measures. However, once the overall capital has been assessed no specific allocation formula is provided or required in order to evaluate the contribution of each risk source on the overall SCR.
The aim of this paper is to provide a closed formula for capital allocation fully coherent with the Solvency II Capital Requirement assessed by means of Standard Formula. The solution proposed permits a top-down approach to assess the allocated SCR among the risks considered in the multilevel aggregation scheme established by Solvency II. Besides, we demonstrate that the allocation formula here proposed is consistent with the Euler’s allocation principle.
\\
\noindent\textbf{Keywords}\\ Capital allocation, Euler principle, Standard Formula, Solvency 2.\\
\noindent\textbf{JEL} G22
\end{abstract}

\clearpage


\section{Introduction}
With the enactment of the new Solvency 2 Directive\cite{EIOPA-Direttiva}, effective starting the 1st of January 2016, a new risk-based solvency standard is introduced in the European insurance market. Part of the Solvency 2 framework is dedicated to the definition of a capital requirement necessary to have sufficient available economic resources to cover both a Minimum Capital Requirement and a Solvency Capital Requirement.
\\
As emanated by article 101 Solvency Capital Requirement 
\textit{shall correspond to the Value-at-Risk of the basic own funds of an insurance or reinsurance undertaking subject to a confidence level of 99.5\% over a one-year period}.
\\
Particularly, the Solvency Capital Requirement is calculated with a standard formula or, possibly, in specific circumstances and subject to the supervisory authorities approval, with partial or full internal models.   In the following, we take into account only the case where Solvency Capital Requirement is calculated by means of standard formula (hereafter referred to as SCR). 
\\
The SCR evaluation follows a modular approach. The global risk which the insurance or reinsurance undertaking is exposed to is divided into risk classes (or modules) each composed of sub-risks (or sub-modules). For each risk class a capital requirement is determined as the aggregation of its sub-risk capital requirement (SCR sub-risks). The capital requirements on risk class level are then aggregated in order to derive the capital requirement for the overall risk. By considering the nature of risks subscribed by an insurance or reinsurance undertaking, this combining of risks that are not fully dependent involves a diversification effect i.e. the overall risk capital related to the combination of sub-risks will be equal or lower than the sum of the capitals for each sub-risk. For more background on the Solvency 2 framework we refer to the official web page of the EIOPA \cite{EIOPA_TS}. Once the overall risk capital for Solvency purposes has been defined, the undertaking has reached the main goal of SII.\\ Nevertheless, in order to analyze capital absorption and/or the economic risk adjusted performance of an insurance portfolio, it is necessary to allocate the diversification effect among each sub-risk and/or sub-portfolio. Diversification forms the foundation of insurance and is the keystone on which important risk management processes rest. To the best of our knowledge Solvency 2 does not provide any specific methodology for capital allocation. The allocation of the SCR, i.e. its calculation net of diversification effect, is a needful procedure to know the real capital absorption of the lines of business and to measure the relative financial performance. Academic researchers have addressed capital allocation for many years proposing several approaches and establishing the principles of coherence through axiomatic definitions for evaluating allocation methods in relation to the specific risk measures (see \cite{Tas1} \cite{Den1} \cite{BucDor}). This line of research has provided significant applications relating to various risk measures assuming different distributions for the underlying risk variable and identifying the Euler’s allocation principle as the highest performing. \\
In this paper we focus on the SCR aggregation formula pointing out its main characteristics and underlying assumptions. Then, we derive an original closed formula to calculate the allocated SCR among the risks considered in the multilevel aggregation scheme established by Solvency II regime, by means of the Euler’s allocation principle. Finally, we compare the result obtained with our formula to other allocation principles and we provide an application for the allocation of overall Basic Solvency Capital Requirement among several Lines of Business. \\

The paper is organized as follows. In section \ref{Par:Theory} we introduce the theoretical framework. In section \ref{SEC:hp FS} we point out properties and remarks about the standard formula provided by EIOPA for SCR calculation. In section \ref{sec:CORPO} we provide an allocation methodology based on the Euler Principle to allocate the Basic Solvency Capital Requirement in the standard formula framework and lastly in section \ref{CaseStudy} we show a numerical application based on the data of a non-life insurance company, comparing our allocation approach with other approaches.

\section{Theoretical framework} \label{Par:Theory}

We consider an insurance or reinsurance undertaking whose portfolio $Q$ of insurance contracts is composed by $q$-homogeneous sub-portfolios. We define a set of random variable $\Gamma$ in the probability space $[\Omega, \Im ,\emph{\textbf{P}}]$.
The profit/loss of the \emph{s-th} ($s=1 ... q)$ sub-portfolio is modeled by means of the generic random variable $X_s \in \Gamma$. The total profit/loss of the company is described with the random variable $X=\sum\limits_{s=1}^q X_s$. 

In order to assess the insurer’s economic capital or the solvency capital requirement for regulatory or internal purposes it is usual to adopt a risk measure on $X$, defined as a functional $\rho$ that maps $X$ to a non-negative real number $\rho(X)$, possibly infinite:

\begin{equation}
\rho(X) : \Gamma \rightarrow \Re
\end{equation}

For an introduction to risk measures, see, for example, Albrecht \cite{Alb1}, Denuit et al. \cite{Denuit1}, Panjer \cite{Panjer2}, McNeil et al. \cite{McNeil1}.
The capital requirement for solvency purpose representing the extra cash that has to be added to expected losses $E(X)$ to cover the unexpected losses is defined as:
\begin{equation}
EC:\pi(X)=\rho(X)-E(X) 
\end{equation}
Several desirable properties for risk measures have been proposed in the literature: see, for example, Denuit et al. \cite{Denuit2}.\\
Once the total economic capital is defined, we are interested in the process of allocating $EC$ across the $q$-sub-portfolios also known as the capital allocation problem: 
\begin{equation}
\pi(X)=\sum\limits_{s=1}^q \pi(X_s|X)
\end{equation}
where, from an economic point of view, $\pi(X_s | X)$ $(s=1, ... , q)$ is the risk contribution net of diversification effect of the $q$-sub-portfolios \cite{Dhaene1}. 
Note that the risk variables $X_s$ are usually dependent so there exist a diversification effect implied in the calculation of the capital requirement $\pi(X)$ such that:
\begin{equation}
\pi(X)=\sum\limits_{s=1}^q \pi(X_s|X) \leq \sum\limits_{s=1}^q \pi(X_s)
\end{equation}
The overall diversification effect is simply measured as:
\begin{equation}
DE=\sum\limits_{s=1}^q \pi(X_s)-\pi(X)
\end{equation}
In the following we summarize certain main risk measure properties as well as capital allocation principle properties useful for our further investigation.

\subsection{Coherent risk measure}
The most important risk measure properties were introduced by Artzner (1999) \cite{Art1} and \cite{Art2} who defines the coherence of a risk measure by means of the following axiom:

\begin{definition} \label{ArtzDeF}
A risk measure $\pi$ is considered coherent if satisfies the following property:
\begin{itemize}
\item \textbf{Translation invariance}: for a riskless deterministic portfolio $L$ with fixed return $\alpha$ and for all $X \in \Gamma$ we have $\pi(X+L)=\pi(X)-\alpha$
\item \textbf{Subadditivity}: for all $(X_1,X_2) \in \Gamma$ we have $\pi(X_1 + X_2) \leq \pi(X_1) + \pi(X_2)$
\item \textbf{Positive Homogeneity}: for all $\lambda >0$ and all $X \in \Gamma$, $\pi(\lambda X) = \lambda \pi(X)$
\item \textbf{Monotonicity}: for all $X,Y \in \Gamma$ with $X \leq Y$, we have $\pi(X) \leq \pi(Y)$
\end{itemize}
\end{definition}

\subsection{Coherent allocation of risk capital}

Denault (2001) \cite{Den1} extends the concept of coherence to the allocation principle establishing a set of definitions and axioms. 
Considering a set $Q=\{1, 2,…, q\}$ of all portfolios of the undertaking and a coherent risk measure $\pi$, the set $A$ of risk capital allocation problems is represented by the pairs $(Q,\pi)$. The following definition holds:
\begin{definition} \label{DEF:AllocationPrinciple}
An allocation principle is a function $\Pi: A \rightarrow \Re ^q$ that maps each allocation problem $(Q,\pi)$ into a unique allocation:\\
\begin{equation}
\Pi: (Q,\pi)\mapsto 
\begin{bmatrix}
    \Pi_1(Q,\pi) \\       
    \vdots  \\
    \Pi_q(Q,\pi)
\end{bmatrix} 
=
\begin{bmatrix}
    K_1 \\       
    \vdots  \\
    K_q
\end{bmatrix} 
\end{equation}
\\
\\
such that $\pi(X)=\sum\limits_{s=1}^q K_s$ (\textbf{\textsl{Full allocation property}})\\
where, followig the notation introduced by Tasche \cite{Tas1}, $K_s=\pi(X_s|X)$ is the allocated risk measure for the sub-portfolio $s-th$.\\
\end{definition}

\begin{definition}
An allocation principle $\Pi$ is coherent if, for every allocation problem, the following three properties are satisfied:

\begin{enumerate}
\item \textbf{No Undercut}

\[
\forall M \subseteq Q, \qquad \sum_{s \in M} \pi(X_s) \leq \pi(\sum_{s \in M} X_s)
\]

\item \textbf{Symmetry}: if by joining any subset $M \subseteq Q \ {i,j}$, portfolios \emph{i} and \emph{j} both make the same contribution to the risk capital, then $\pi(X_i|X)=\pi(X_j|X)$.

\item \textbf{Riskless allocation}: for a riskless deterministic portfolio $L$ with fixed return $\alpha$ we have that 
\[
\pi(L) = -\alpha
\]
\end{enumerate}
\end{definition}

\subsection{Main allocation principle} \label{main_allocation_principle}

There are several allocation principle commonly used in practice (see \cite{McNeil1}). They imply different sets of assumptions and their applicability depends upon the circumstances. Among these, we introduce the \emph{Euler}, \emph{haircut}, \emph{marginal}, \emph{covariance} and \emph{market driven} allocation principles. 

\subsubsection{Euler allocation principle}

The Euler allocation principle derives from the well-known  \emph{Euler's homogeneous function theorem} applied to a risk measure. The method states that if the risk measure to be allocated a first degree homogeneous function, then it is possible to represent it as follows:

\begin{equation}
\pi(X) = \pi \left(\sum_{s=1}^q X_s\right) = \sum_{s=1}^q \pi(X_s) \cdot \frac{\partial \pi(X)}{\partial \pi(X_s)} = \sum_{s=1}^q \pi(X_s|X)  \qquad s\in Q
\end{equation}

In this way, the value of the reference risk measure, is represented as the sum of additive components, each representing the value of the risk measure for the variable \emph{i-th} net of diversification.

Euler Allocation Principle is appealing for its economic meaning: to give more weight to risk where the overall capital is more sensitive.

\subsubsection{Haircut allocation principle}

A straightforward way to allocate capital is based on assumption of proportionality between allocated and unallocated capital requirement, as for:

\begin{equation}
    \pi \left(X_s | X \right) = \pi \left(X \right) \cdot \frac{\pi \left(X_s \right)}{\sum_{s=1}^q \pi \left(X_s \right)}
\end{equation}

This allocation principle is very easy to compute but it does not take into account the correlation among risks so it can only be used where there is no correlation effect. 

\subsubsection{Marginal allocation principle}

This principle allows to allocate capital by considering the stand alone risk contributions of the several sub-risks to the total risk capital:

\begin{equation}
    \pi \left(X_s | X \right) = \pi \left(X \right) \cdot \frac{\pi \left(X \right) - \pi \left(X -X_s \right)}{\sum_{i=1}^q \pi \left(X \right) - \pi \left(X -X_i \right)}
\end{equation}

The principle takes into account the correlation effect implicitly. It provides a numerical approximation for the partial derivative of the risk measure with respect to one specific risk variables. This gives results that take into account the correlation among risks, but the accuracy is lower than Euler principle. Furthermore, its calculation requires a number of iterations equal to the number of sub-portfolios considered.

\subsubsection{Covariance allocation principle}

This principle, since the variability of total risk capital is fully explained by the sub-risks, starts from the following variance decomposition formula:

\begin{equation*}
VAR(X) = \sum_{s=1}^q COV(X_s , X)    
\end{equation*}

so that

\begin{equation}
     \pi \left(X_s | X \right) = \pi \left(X \right) \cdot \frac{COV \left(X_s , X \right)}{VAR \left(X \right)}
\end{equation}

\subsubsection{Market Driven allocation principle}

This principle consists in a simple proportional rule among the allocated capital and a variable assumed as risk driver:

\begin{equation}
    \pi \left(X_s | X \right) = \pi \left(X \right) \cdot \frac{RD_s}{\sum_{s=1}^q RD_s}
\end{equation}

where $RD_s$ is the $s-th$ risk driver.

\subsection{On the coherence of Euler allocation principle}
The Euler’s allocation principle described in the previous subsection, is one of the most popular allocation methods proposed in literature. This is due to its suitable properties. In this sense, a very important contribution is that of Buch et G. Dorfleitner (2008) \cite{BucDor}. From an axiomatic point of view, they study the relation between the properties of the Euler's allocation principle and those of the risk measure to which the allocation is applied. What they find is summarized in the following proposition.
\begin{proposition}
\label{PROP:Buch_Dorfleitner}

The Euler's allocation principle applied to a coherent risk measure has the properties of "full allocation", "no undercut" and "riskless allocation" so it is coherent with the definition given by Denault (2001) \cite{Den1}.
\end{proposition}

\clearpage
\section{Solvency II standard formula: basic assumptions and general framework} \label{SEC:hp FS}
The Solvency II framework is a regulatory project that imposes (re)insurance undertakings to calculate regulatory solvency capital requirement by means of a risk based methodology \cite{EIOPA-Direttiva}.
From a practical point of view, undertakings SCR may be calculated by means of the standard formula, provided by EIOPA, or via (partial) internal model. Although the latter approach may be better to match the risk profile of the entity, its adoption must be approved by the supervising authority through a rigorous procedure. Otherwise, the Standard Formula approach may be considered the benchmark Supervisor's method and is widely adopted by market participants to calculate their SCR or as a comparative measure with a (partial) internal model.\\
Due to its strategic relevance for insurance market participants, in the following we take into account only the SCR calculated with the standard formula. It considers that the insurance company must compute the overall risk exposure by considering a set of specified risk sources. \\
The risk-based modular approach considered in the Solvency II framework provides that the insurance company has to consider its global risk by dividing it into single components, each one related with its specific source of risk. 
The modular scheme considers $n$ risk modules\footnote{Actually there are six risk modules: market risk, non-life underwriting risk, life underwriting risk, health underwriting risk, default risk, intangible asset risk}. The generic risk module $i-th$ ($i=1, ... , n$) is composed by $m_i$ sub-risks. We use the following notation for all variables that will be defined: the first digit of the subscript identifies the risk module and is from $1$ to $n$, the second one identifies the sub-risk and is from $1$ to $m_i$ (where $i$ identifies the overlying risk module).
\\
\begin{definition}[Standard Formula] The solvency II capital requirement is defined by means of a modular bottom up approach as follows \cite{EIOPA_TS}: 
\label{def:SF}
\begin{enumerate}[I]
 \item $SCR_{ij}$ is the capital requirement referred to the $ij-th$ sub-risk and is calculated by means of a set of specific formulas provided by EIOPA.
 \item $SCR_{i}$ is the capital requirement referred to the $i-th$ risk-module calculated by aggregating the underlying sub-risks:
 
\begin{equation} \label{EQ:SCR_iy}
SCR_{i}=\sqrt{\sum_{x=1}
^{m_i} \sum_{y=1}
^{m_i} SCR_{ix} \cdot SCR_{iy} \cdot \rho_{ix,iy}} 
\end{equation}

where $\rho_{ix,iy}$ represents the linear correlation coefficients. They are provided by EIOPA and are equal for all insurance companies.

 \item $BSCR$ is the capital requirement performed aggregating the underlying risk-modules:
 
\begin{equation} \label{EQ:SCR_i}
BSCR= \sqrt{\sum_{i=1}
^{n} \sum_{w=1}
^{n} SCR_{i} \cdot SCR_{w} \cdot \rho_{i,w}} 
\end{equation}
\\
where $\rho_{i,w}$ represents the linear correlation coefficients. They are provided by EIOPA and are equal for all insurance companies.

\item $SCR$ is the overall capital requirement performed by adding certain other components to the BSCR:
 
\begin{equation} \label{EQ:SCR}
SCR= BSCR + Adj + OP_{risk} 
\end{equation}
\\
where $Adj$ represents the adjustment for deferred taxes and for loss-absorbing capacity of technical provisions and $OP_{risk}$ is the capital requirement for operational risk.
\end{enumerate}
\end{definition}

In this paper we consider only the BSCR excluding Adjustments and Operational Risk.  This because their effect is measurable after calculating   the SCR and their allocation depends not on the aggregation scheme but on particular considerations made by the Company. Furthermore, the adjustment for deferred taxes calculation (whose relevance can be very high) requires the sub-risk allocation of the BSCR as an input.

\subsection{Standard Formula properties and remarks} \label{Par:SF properties and remarks}

The above reported aggregation formulas (Equations \ref{EQ:SCR_iy} and \ref{EQ:SCR_i}) call for further considerations, and these may be illustrated as follows:
\begin{enumerate}[i]
\item  As stated by EIOPA the overall SCR shall correspond to a specific risk measure, the Value-at-Risk ($VaR$), subject to a confidence level of 99.5\% over a one-year period so, in the intention of the Supervisor, it seems acceptable to write $SCR\approx VaR_{99.5\%}$.
\item In \cite{EIOPA-assumption} EIOPA has specified that the correlation matrices used for the aggregation of sub-risks (Eq. \ref{EQ:SCR_iy}) and risk modules (Eq. \ref{EQ:SCR_i}) respectively, are estimated to minimize the aggregation error through the following formulation:
\begin{equation}\label{minrho}
\left|\min_{\rho} VaR(X+Y)^2 - VaR(X)^2 - VaR(X)^2 -2 \rho VaR(X) VaR(Y)\right|
\end{equation}
where $X$ and $Y$ are random variables that represent two different risks.\\
\item Each $SCR_{ix}$ is calculated by means of specific methodologies stated by EIOPA. The general principle for the calculation of a single sub-risk capital requirement $SCR_{ix}$ is to apply a set of shocks to the risk drivers and calculate the impact on the value of the assets and liabilities. The calibration objective - i.e. the calibration using Value at Risk subject to a confidence level of 99.5\% over a one-year period - is extended to each individual risk module and sub-risk.\\
\end{enumerate}

\begin{remark}
As well known, the structure with a square root of a quadratic expression and the use of correlation matrices produce a correct aggregation of quantiles in case of any centered elliptical distribution, such as the (multivariate) normal distribution\footnote{In case the expected values of the marginal distributions are zero. This simplifying assumption is made in the standard formula which intends to quantify unexpected losses.}.
\end{remark}

\begin{remark}
In the SCR calculation EIOPA does not put forward assumptions for the distribution of the losses of each risk class and/or sub-risk, but the underlying assumption of linear correlation and elliptic distribution are implicit and necessary for the correctness of the aggregation formulas.
These assumptions are very strong because, as well known, in insurance problems the dependence among probability distributions is not linear just as tail dependencies and the shape of the marginal distributions are usually not skewed. As stated by Sandstrom (2007) \cite{Sandstrom1} for skewed distribution the normal approximation can imply an incorrect estimation of the SCR and he proposes a method to transform, via Cornish-Fisher expansion, the quantile distribution from a skewed into a standard normal distribution.\\
\end{remark}

\begin{remark}
Leaving aside the well known criticism on VaR, on the (implicit) elliptical distribution assumption of each risk module or sub-risk and on the use of linear correlation among risks, another relevant issue to the bottom-up aggregation approach proposed by EIOPA is that it does not represent a `genuine' bottom-up approach to risk aggregation. By nesting (\ref{EQ:SCR_iy}) in (\ref{EQ:SCR_i}) as follows:

\begin{equation}
\label{EQ:SCR_nested}
BSCR =  \sqrt{\sum_{i=1}^{n} \sum_{j=1}^{n} \left[\sqrt{\mathbf{SCR_{i\bullet}^{\mathrm{T}}} \cdot \mathbf{P_i} \cdot \mathbf{SCR_{i\bullet}}}\right] \cdot \left[\sqrt{\mathbf{SCR_{j\bullet}^{\mathrm{T}}} \cdot \mathbf{P_j} \cdot \mathbf{SCR_{j\bullet}}}\right] 
\cdot \rho_{i,j}}
\end{equation}

Equation (\ref{EQ:SCR_nested}) is in general inconsistent with any multivariate distribution of risks. As observed by Filipovic (\cite{Filip1}) a genuine bottom-up model uses a full base correlation matrix $\textbf{B}:M \times M \rightarrow\Re$ (where $M=m_1+m_2+...+m_n$) that aggregates all risk types, across risk classes, together:
\begin{equation} \label{EQ:SCR_genuine}
SCR=\sqrt{\textbf{A}^{\mathrm{T}} \cdot \mathbf{B} \cdot \mathbf{A}}\\
\end{equation}

where, $\mathbf{A}=[\mathbf{SCR_{i\bullet}},  ... , \mathbf{SCR_{n\bullet}}]$  is the vector of all sub-risk capital requirement vectors.\\
Nevertheless, the only available information about correlation is contained in each matrix $\mathbf{P_i}, i=1,2,...,n$ but is limited to the correlation coefficients among sub-risks referred to the same risk modules. The missing correlation coefficients in $\mathbf{B}$ are referred to sub-risks belonging to different risk modules (e.g. equity risk and lapse risk) whose estimate is an arduous task.
\end{remark}

Finally, the risk aggregation bottom-up approach provided by EIOPA has the following properties and shortcomings:

\begin{itemize}
\item the overall SCR is based on a $VaR$ risk measure so it involves all the coherent risk measure properties\footnote{see Artzener et al\cite{Art1}} excluding sub-additivity; 
\item the implicit elliptical distribution assumption underlying Equations \ref{EQ:SCR_iy} and \ref{EQ:SCR_i}) involves the sub-additivity property;
\item the nested aggregation formula (\ref{EQ:SCR_nested}) is homogeneous of the first degree;
\item the two-step aggregation formula proposed by EIOPA is inconsistent with any multivariate probability distribution and does not represent a genuine bottom-up approach as stated by Filipovic \cite{Filip1}.\\
\end{itemize}

The SCR computed using the Standard Formula should be interpreted as a risk indicator that, given the formal inconsistencies of the aggregation approach based on a unique standardized methodology permits a proxy of the VaR for the unexpected loss only ideally. Notwithstanding the above mentioned issues, it is suitable to represent the overall solvency condition of an insurance undertaking because its value is coherent with the nature of risks assumed by the Company and, moreover, it increases (or decreases) according to higher (or lower) risk assumed.\\ \\

\section{Capital allocation of SCR under Solvency II Standard Formula} \label{sec:CORPO}

\textbf Notwithstanding the above mentioned limits, the Solvency II aggregation Standard Formula is largely adopted by (re)insurance undertakings in EU countries to determine the overall risk capital. Due to the implicit sub-additivity property,  it involves a diversification effect that reduces the $SCR$ in each aggregation step. 
Once the diversification effect is determined we want to know what amount the undertaking allocates to each risk-module or sub-risk, in order to know the real capital absorption of each risk or to measure the financial performance of each line of business or product. 
So, based on the properties of the Standard Formula previously introduced, in the following we show how to assign the $BSCR$, net of diversification effect, among sub-risks and coherently with the aggregation formulas using a top-down approach as follows:
\begin{enumerate}
\item \textbf{risk-allocation}: allocate the $BSCR$  among each $i-th$ risk module to define the relative allocated capital $SCR_i^A = \pi (X_{i}|X)$, so that the full allocation principle is respected i.e. $BSCR=\sum_{i=1}^{n} SCR_i^A$;
\item \textbf{subrisk-allocation} 
\begin{enumerate}
\item allocate the $i-th$ risk module solvency capital requirement $SCR_i = \pi (X_i)$ among each $iy-th (iy=i1,...,im_i)$ sub-risk to define the relative allocated capital $SCR_{iy}^{Ai} = \pi (X_{iy}|X_i)$, so that the full allocation principle is respected i.e. $SCR_{i}=\sum_{y=1}^{m_i} SCR_{iy}^{Ai}$;
\item allocate $BSCR$  among each $iy-th$ sub-risk to define the relative allocated capital $SCR_{iy}^A = \pi (X_{iy}|X)$, so that the full allocation principle is respected i.e. $BSCR=\sum_{i=1}^{n} \sum_{y=1}^{m_i} SCR_{iy}^A$;
\end{enumerate}
\item \textbf{LoB-allocation}: allocate the allocated $SCR_{iy}^A$ among sub-portfolios or Lines of Business.
\end{enumerate}

In the following we demonstrate that the first two allocation steps above reported are obtained by closed formulas while the allocation of sub-risk capital requirement among Lines of Business is extendible to closed formulas only where a square root aggregation formula is used for its calculation; e.g. to allocate the Premium-Reserve Risk included in the Non-Life and Health underwriting risk among the different Lines of Business defined by EIOPA.\\
Starting from equation (\ref{EQ:SCR_i}) it is possible to obtain an explicit expression of the allocated capital in each $i-th$ risk-module $SCR_{i}^A$ as follows:
\begin{equation}
\begin{split} 
  BSCR &= \sqrt{\sum_{i=1}^n \sum_{j=1}^n SCR_i \cdot SCR_j \cdot \rho_{i,j}} = \\ \\
  &= \frac{\sum_{i=1}^n \sum_{j=1}^n SCR_i \cdot SCR_j \cdot \rho_{i,j}}{BSCR} = \\ \\
  &= \sum_{i=1}^n  SCR_i \cdot \frac{\sum_{j=1}^n SCR_j \cdot \rho_{i,j}}{BSCR}
\end{split}
\end{equation}
\\
The \emph{i-th} net of diversification component is:
\\
\begin{equation} \label{EQ:SCR_alloc}
SCR_i^A = SCR_i \cdot \frac{\sum_{j=1}^n  SCR_j \cdot \rho_{i,j}}{BSCR}
\end{equation}
\\ \\
The solution stated in equation (\ref{EQ:SCR_alloc}) is unique and respects the Full Allocation property. Moreover, as we demonstrate in the following theorem it is fully compliant with the Euler allocation principle:

\begin{theorem}[$BSCR$ allocation on risk module] \label{RiskModulesTheorem}
In the case of the Solvency II Standard Formula, the Euler allocation of the $BSCR$ among the underlying risk modules is uniquely determined as:
\begin{equation}\label{eq:SCR_allocated_Eul1} 
  BSCR= \sum\limits_{i=1}^n SCR_i^A= \sum\limits_{i=1}^n SCR_i \cdot \frac{\sum\limits_{j=1}^n  SCR_j \cdot \rho_{i,j}}{BSCR}
\end{equation}

where $SCR_i^A$ is the amount of the $BSCR$ allocated on \emph{i-th} risk module.

\end{theorem}

\begin{proof}

From (\ref{EQ:SCR_i}) we have that $BSCR=f(SCR_1,...,SCR_n)$ is a first degree homogeneous function, so from the Euler's homogeneous function theorem we obtain:
\begin{equation}\label{eq:SCR_Eul1} 
  BSCR= \sum\limits_{i=1}^n SCR_i \cdot \frac{\partial BSCR}{\partial SCR_i} 
\end{equation}
where the partial derivative of $BSCR$ respect to $SCR_i$ is:
\begin{equation}
\frac{\partial BSCR}{\partial SCR_i}= \frac{\sum\limits_{j=1}^n  SCR_j \cdot \rho_{i,j}}{BSCR} 
\end{equation}
 
so it results:
\begin{equation}
  SCR_i^A= SCR_i \cdot \frac{\sum\limits_{j=1}^n  SCR_j \cdot \rho_{i,j}}{BSCR}
\end{equation}
\end{proof}
It is useful to define the \textit{Allocation Ratio}, $0 \leq AR_i \leq 1$, as: 
\begin{equation}
AR_i = \frac{\partial BSCR}{\partial SCR_i}=\frac{\sum\limits_{j=1}^n  SCR_j \cdot \rho_{i,j}}{BSCR}
\end{equation}
\\
As previously stated, starting from (\ref{EQ:SCR_iy}) we find out similar results for the $SCR_i$ since it is a first degree homogeneous function and we can obtain the risk module capital allocation on each related sub-risk.
\begin{theorem}[$SCR_i$ allocation on sub-risk] \label{SubRiskTheorem}
In the case of the Solvency II Standard Formula, the Euler allocation of the $i-th$ risk module $SCR_i$ among the underlying sub-risks is uniquely determined as:
\begin{equation}\label{eq:SCR_i_Eul1} 
SCR_i=\sum\limits_{y=1}^{m_i} SCR_{iy}^{Ai} = \sum\limits_{y=1}^{m_i} SCR_{iy} \cdot \frac{\sum_{w=1}^{m_i} SCR_{iw} \cdot \rho_{iy,iw}}{SCR_i}
\end{equation}
\\
where $SCR_{iy}^{Ai}$ is the amount of $SCR_i$ allocated on \emph{y-th} sub risk and $0 \leq AR_{iy} \leq 1$ is the relative \textit{Allocation Ratio}: 
\begin{equation}\label{eq:SCR_iy_Eul1} 
AR_{iy} = \frac{\partial SCR_i}{\partial SCR_{iy}}=\frac{\sum_{w=1}^{m_i} SCR_{iw} \cdot \rho_{iy,iw}}{SCR_i}
\end{equation}
\end{theorem}

\begin{corollary} \label {corollario}
From theorems \ref{RiskModulesTheorem} and \ref{SubRiskTheorem} we have:
\begin{equation}\label{disugualianza_teoremi}
BSCR= \sum\limits_{i=1}^n SCR_i^A \leq \sum\limits_{i=1}^n SCR_i=\sum\limits_{i=1}^n  \sum\limits_{y=1}^{m_i} SCR_{iy}^{Ai}
\end{equation} 
\end{corollary}

As a consequence it is necessary to find an alternative solution to allocate the $BSCR$ on each sub risk, as proposed in the following theorem:. \\
\begin{theorem}[$BSCR$ allocation on sub-risk] \label{SCRsubRiskTheorem}
In the case of the Solvency II Standard Formula, the Euler allocation of the BSCR among underlying sub-risks is uniquely determined as:

\begin{equation} \label{eq:SCR_Eul_2} 
\begin{split} 
BSCR &= \sum\limits_{i=1}^n \sum\limits_{y=1}^{m_i} SCR_{iy}^A=\\ \\
& = \sum\limits_{i=1}^n \sum\limits_{y=1}^{m_i}  SCR_{iy} \cdot AR_{iy} \cdot AR_i=\sum\limits_{i=1}^n \sum\limits_{y=1}^{m_i} SCR_{iy}^{Ai} \cdot AR_i
\end{split} 
\end{equation}
where the variable $SCR_{iy}^A$ is the amount of the overall $BSCR$ allocated on \emph{y-th} sub risk.
\end{theorem}
\begin{proof}
From (\ref{EQ:SCR_nested}) we have that $BSCR=f(SCR_{11},..,SCR_{1m_1},..,SCR_{n1},..SCR_{nm_n})$ is a first degree homogeneous function, so from the Euler's homogeneous functions theorem we obtain:
\begin{equation}
BSCR=\sum\limits_{i=1}^n \sum\limits_{y=1}^{m_i} SCR_{iy}^A=\sum\limits_{i=1}^n \sum\limits_{y=1}^{m_i} SCR_{iy} \cdot \frac{\partial BSCR}{\partial SCR_{iy}}
\end{equation}

By using elementary algebra we know:

\begin{equation} \begin{split}
\frac{\partial BSCR}{\partial SCR_{iy}}= \frac{\partial BSCR}{\partial SCR_{i}} \cdot \frac{\partial SCR_i}{\partial SCR_{iy}}=AR_i \cdot AR_{iy}
\end{split}\end{equation}

thus:

\begin{equation}
  SCR_{iy}^A= SCR_{iy} \cdot \frac{\sum\limits_{w=1}^{m_i}  SCR_{iw} \cdot \rho_{iy,iw}}{SCR_i} \cdot AR_i
\end{equation}
\end{proof}

For practical use, note that:

 \begin{equation}
AR_i = \frac{\partial BSCR}{\partial SCR_i}= \frac{\sum\limits_{j=1}^n  SCR_j \cdot \rho_{i,j}}{BSCR} = \frac{SCR_i^A}{SCR_i}
\end{equation}

By summarizing the above reported theorems allows us to express the following relationship:
\begin{equation}\label{eq:SCR_Eul3}
\begin{split}
BSCR & =\sum\limits_{i=1}^{n} SCR_{i}^A=\sum\limits_{i=1}^{n} \sum\limits_{y=1}^{m_i} SCR_{iy}^A= \\ \\ 
& =\sum\limits_{i=1}^{n} \sum\limits_{y=1}^{m_i} SCR_{iy} \cdot \frac{\sum\limits_{w=1}^{m_i}  SCR_{iw} \cdot \rho_{iy,iw}}{SCR_i} \cdot AR_i
\end{split}
\end{equation}

Theorems \ref{RiskModulesTheorem}, \ref{SubRiskTheorem} and \ref{SCRsubRiskTheorem} provide closed formulas for the capital requirement allocation among risk modules and sub-risks based on the Euler allocation principle. In the following, we refer to this result as \emph{Standard Formula Euler Principle} ($SFEP$).
\\ \\
As a summary, in this section we provide the three results as follows:

\paragraph{BSCR Allocation among risk modules}
\hspace{0pt} \\

From theorem \ref{RiskModulesTheorem}  

\begin{equation} \label{EQ:risk_allocation_REF}
  SCR_i^A= SCR_i \cdot \frac{\sum\limits_{w=1}^n  SCR_w \cdot \rho_{i,w}}{BSCR}
\end{equation}

where $SCR_i^A$ is the amount of SCR allocated on $i-\emph{th}$ risk modules.

\paragraph{BSCR Allocation among sub-risk} 
\hspace{0pt} \\

From theorem \ref{SCRsubRiskTheorem}  

\begin{equation} \label{EQ:subrisk_allocation_REF}
  SCR_{iy}^A=  SCR_{iy} \cdot \frac{\sum\limits_{w=1}^{m_i}  SCR_{iw} \cdot \rho_{iw,iy}}{SCR_i} \cdot AR_i
\end{equation}

where $SCR_{iy}^A$ is the amount of SCR allocated on $iy-\emph{th}$ risk modules.

\paragraph{A generalization for BSCR allocation among sub-risk based on a r-level square root aggregation scheme} 
\hspace{0pt} \\

More in general, when a square root aggregation formula as the one proposed by EIOPA is utilized to aggregate $r$ levels, the top-down allocation formula in the lower level is:
\begin{equation} \label{EQ:r-level}
\begin{split}
BSCR & =\sum\limits_{l_1,l_2,...,l_r} SCR_{l_1,l_2,...,l_r}^{A} = \\ \\
& = \sum\limits_{l_1,l_2,...,l_r} SCR_{l_1,l_2,...,l_r} \cdot \frac{\sum\limits_{h=1}^{m_{l_r}} SCR_{l_1,l_2,...,,l_{r-1}l_h} \cdot \rho_{r,h}^{(l_1,l_2,...,l_{r-1})}}{SCR_{l_1,l_2,...,l_{r-1}}} \cdot \prod_{s=1}^{r-1} AR_s
\end{split}
\end{equation}

where the \textit{Allocation ratio} at $s-th$ level is:

\begin{equation}
AR_s = \frac{\partial SCR_{l_1,...,l_{s-1}}}{\partial SCR_{l_1,...,l_s}}
\end{equation}

\section{Case Study} \label{CaseStudy}

In this section we provide a first simple example to perform a comparison between the $SFEP$ and the other allocation methodologies introduced in \ref{main_allocation_principle}. Our aim is to assess the ineffectiveness of the latter methods to allocate SCR when the square root aggregation formula in used. Furthermore, we show an application of the proposed method for a full allocation of $SCR$ on the single Line of Business (LOB),  as defined by EIOPA \cite{EIOPA_Annex_K}, based on a true data set provided by an anonymous Non-Life insurance undertaking.
\\
\\
\subsection{Comparison between allocation principles}
For the $SCR$ calculation and the subsequent allocation we consider a two-step aggregation scheme based on $n=3$ risk-modules composed by $m_i=2$ sub-risks, with $i=1,2,3$. The square root aggregation formulas (\ref{EQ:SCR_iy}) and (\ref{EQ:SCR_i}) are used to assess the $SCR_i, i=1,2,3$ and the overall $SCR$, respectively. The capital requirement for each sub-risk expressed in m.u. is:
\begin{table}[!ht]
\centering
\caption{Sub-risk Capital Requirement}
\begin{tabular}{l r r r}
\hline
Risk -Module & $y=1$ & $y=2$ & Tot \\ 	
\hline
$SCR_{1_y}$  & 60   & 70   & 130\\ 
$SCR_{2_y}$  & 110   & 130 & 240\\ 
$SCR_{3_y}$  & 45   & 70  & 115\\  
\hline
Tot  & -   & -  & 485\\  
\hline
\end{tabular}
\end {table}
\\
Assuming the following correlation matrices:
\\
\begin{equation}
\mathbf{P_i}=
\begin{bmatrix}
    1 &  0\\  
    0 & 1 \\
\end{bmatrix}
\end{equation}
\\
\begin{equation}
\mathbf{P}=\begin{bmatrix}
    1 & 0.5 & 0.5 \\  
    0.5 & 1 & 0.5\\
		0.5 & 0.5 & 1\\
\end{bmatrix}
\end{equation}

By applying the square-root aggregation formula (\ref{EQ:SCR_iy}) we get:
\begin{table}[!ht]
\centering
\caption{Risk Module Capital Requirement}
\begin{tabular}{l r r}
\hline
Aggregation Level &  $SCR$ & $DE$\\ 	
\hline
$SCR_1$    & 112.69  & 17.31\\
$SCR_2$ &  208.09 & 31.91\\  
$SCR_3$  &  100.37 & 14.63\\  
\hline
Tot  &  421.16 & 63.84\\  
\hline
\end{tabular}
\end{table} 
\\
As a result of the aggregation phase using (\ref{EQ:SCR_i}) we get an overall $SCR$ of 257.05 m.u. with a diversification effect among risk-modules of 164.10 m.u.. The overall diversification effect in the two-step aggregation method is 227.95 m.u. i.e. a decreases of about 53\% of the sum of the capital requirement of each sub-risk.\\
In order to spread the overall $SCR$ on each sub-risk we use the $SFEP$ provided in this paper compared to the Marginal Principle and the Haircut Principle introduced in \ref{main_allocation_principle}. We have not compared it with the Covariance Principle as well because its application requires to know both variance and covariance among each sub-risk: we are not able to know these due to general inconsistency of the aggregation scheme of the standard formula with any multivariate probability distribution of risks\footnote{as observed by Filipovic (\cite{Filip1} and remarked in section \ref{Par:SF properties and remarks}.}.

\begin{table}[!ht]
\centering
\caption{Allocation Principle comparison}
\label{table:confronti}
\begin{tabular}{l r r r r r}
\hline
Level & $SFEP$ & Marginal & Haircut & Marginal vs $SFEP$& Haircut vs $SFEP$\\ 	
  & & &  &  in \% &  in \% \\
\hline
$SCR_1$  & 49.41   & 43.84  & 68.78 &-11.27 & 39.22\\
$SCR_2$ & 168.45  & 178.43 & 127.00 &6.16 & -24.60\\
$SCR_3$   & 39.19   & 34.38  & 61.26 &-12.27 & 56.30\\
\hline
Tot & 257.05 & 257.05 & 257.05 & 0.00& 0.00\\
\hline
$SCR_{1_1}$    & 22.17   & 21.62  & 31.80 &-2.49 &43.41\\
$SCR_{1_2}$     & 27.23   & 24.77  & 37.10 &-9.04 &36.24\\
$SCR_{2_1}$    & 74.89   & 80.04  & 58.30 &6.88 &-22.15\\
$SCR_{2_2}$    & 93.56   & 94.77  & 68.90 &1.29 &-26.36\\
$SCR_{3_1}$     & 14.01   & 14.50  & 23.85 &3.56 &70.30\\
$SCR_{3_2}$    & 25.19   & 21.35  & 37.10 &-15.25 &47.28\\
\hline
Tot & 257.05 & 257.05 & 257.05 & 0.00 & 0.00\\
\hline
\end{tabular}
\end{table}

The outcomes above reported show that:
\begin{itemize}
    \item the Haircut principle produces a capital allocation strongly different from $SFEP$ and is not informative from a risk management point of view as it fully respects the initial capital requirements of each sub-risk and does not take into account the correlation among risks;
    
    \item the Marginal principle provides a numerical approximation for the partial derivative of the total $SCR$ related to each specific sub-risk. This allows to obtain a proxy of the allocated capital coherent with the exact allocation produced with the $SFEP$ in terms of sign. However, it produces a larger effect in terms of absolute values.
  
\end{itemize}

\subsection{BSCR Allocation among Lines of Business for a Non-life insurance} 
In section \ref{sec:CORPO} we describe the procedure to be followed to allocate BSCR among risks, sub risks and LoB. In this section we apply the  methodology described in section \ref{sec:CORPO} focusing our attention on the allocation of the so-called \textit{Non-Life Underwriting Risk}  risk module on its sub risks, i.e. Premium \& Reserve, Lapse and Catastrophe. The aim is to perform an accurate capital allocation among LoB considering that Premium \& Reserve and CAT sub-risk $SCR$ are obtained, under Standard Formula, with a square root aggregation formula; in these cases, a three-level aggregation scheme is adopted by EIOPA. For other risk modules the allocation formula here proposed is applicable for the sub risks allocation. In any case, a market driven or other allocation principle may be adopted for LOB allocation.

\subsubsection{Data set and $SCR$ calculation.}
The $BSCR$ is obtained by the aggregation of the following risk modules : 1 - Market, 2 - Default, 3 - Life Underwriting, 4 - Health Underwriting and 5 - Non-Life Underwriting. Without loss of generality, we are not considering the Intangible risk.\\
The first step we perform to calculate $BSCR$ is the calculation of $SCR_5$ (Non-life underwriting risk module) by aggregating its sub-risks ($SCR_{51}$ for Premium \& Reserve Risk, $SCR_{52}$ for Lapse Risk and $SCR_{53}$ for CAT Risk). Thus, as a first step, we start from the Premium \& Reserve Risk whose capital requirement is computed by means of a square root function of the volume measures of Premium and Reserve risks similar to (\ref{EQ:SCR_iy}). Although not explicitly stated in \cite{EIOPA_TS}, the Premium \& Reserve risk for each LoB can be alternatively obtained by separately computing $SCR_{premium}$ and $SCR_{reserve}$ and then aggregating with a linear correlation coefficient of 0.5. It follows that:

\begin{table}[!ht]
\centering
\caption{Best Estimate Liability and Premium \& Reserve Risk SCR}
\label{table:BEL}
\begin{tabular}{l l r r r}
\hline
k & LoB Name & $SCR_{premium}$ & $SCR_{reserve}$ & $SCR_{51k}$ \\ 	
\hline
 1 	& Motor vehicle liability           & 	 673,397 	& 	 3,269,802 	& 	 3,653,347   \\ 
 2  & Other motor 	                    & 	 1,056,640 	& 	 2,550,459 	& 	 3,211,891   \\ 
 3 	& Marine, aviation and transport    & 	 1,475,581 	& 	 1,730,753 	& 	 2,779,696   \\ 
 4 	& Fire and other damage to property & 	 646,519 	& 	 1,702,827 	& 	 2,102,026   \\ 
 5 	& General liability                 & 	 840,929 	& 	 3,090,863 	& 	 3,586,055   \\ 
 6 	& Credit                            & 	 542,467 	& 	 681,076 	& 	 1,061,883   \\ 
 7 	& Legal expenses                    & 	 184,146 	& 	 2,545,219 	& 	 2,642,109   \\ 
 8 	& Assistance                        & 	 1,306,716 	& 	 491,145 	& 	 1,609,509   \\ 
 9 	& Miscellaneous financial loss      & 	 1,901,405 	& 	 5,677,832 	& 	 6,830,006   \\ 

\hline
$\Sigma$ &    &  30,625,400     &     61,570,890    &   27,476,524             \\
\hline
\end{tabular}   
\end{table}
As reported in Table \ref{table:BEL}, the company has a good risk diversification among LoBs and the net of diversification ($DE$) capital requirement is:\\
$SCR_{51}=19,490,560$ and $DE_{51}=27,476,524-19,490,560=7,985,964$.\\ \\
The CAT Risk ($SCR_{53}$) requires the aggregation of two main sub risks, Man Made and Natural, and within the latter a distinction between natural events such as Earthquake, Flood, etc. (for further details see \cite{EIOPA_TS}). The $SCR_{53}$ is obtained by using a double level aggregation square root formula where correlation among risks is generally null. In order to resume the outcomes we limit ourselves to report the value obtained:\\
$SCR_{53}=10,248,826$ and $DE_{53}=20,087,825-10,248,826= 9,838,999$.\\ \\
The Lapse Risk ($SCR_{52}$) requires a scenario based approach depending on two shocks and it does not require an aggregation formula, so we get the information from our dataset where:\\
$SCR_{52}=552,645$. \\ \\
The second aggregation step consists in the application of (\ref{EQ:SCR_iy}) with $i=5$ and $m_i=3$; so we obtain:\\
 $SCR_{5}= 24,188,911$ and $DE_{5}=31,135,851-24,188,911=6,103,119$.\\ \\
The third and last aggregation level is obtained by applying (\ref{EQ:SCR_i}) to the $n=5$ risk modules:\\
$BSCR=29,647,059$ and $DE_{BSCR}=6,218,424$. \\

\subsubsection{BSCR allocation among risk modules} 

After calculating the $BSCR$ as previously stated, the first step for a top-down allocation procedure is to allocate the $BSCR$ among the 5 risk modules. For the allocation we use the (\ref{EQ:risk_allocation_REF}):
\begin{equation}
  SCR_i^A= SCR_i \cdot \frac{\sum\limits_{w=1}^5  SCR_w \cdot \rho_{i,w}}{BSCR}
\end{equation}

\begin{table}[!ht]
\centering
\caption{Risk module Allocation 1}
\label{table:Risk_module_SCR 1}
\begin{tabular}{l l r r r}
\hline
Risk Module &	$i$ & $SCR_i$ &	$SCR_i^A$ &	$AR_i$\\
\hline
Market                 & 1  & 6,112,345     & 2,793,738     & 46\% \\
Default                & 2  & 5,564,226     & 3,601,015     & 65\% \\
Life Underwriting	   & 3  &       0       &     0         &    -    \\
Health Underwriting	   & 4  &   0           &       0       &  -      \\
Non-Life Underwriting  & 5  & 24,188,911    & 23,252,305    & 96\% \\
\hline
Total                  &    & 35,865,424    & 29,647,059 \\     
\hline
\end{tabular}
\end{table}

\begin{table}[!ht]
\centering
\caption{Risk module Allocation 2}
\label{table:Risk_module_SCR 2}
\begin{tabular}{l r}
\hline
$\sum_i SCR_i$      & 35,865,482    \\
\hline
Diversification     & 6,218,424     \\
\hline
BSCR                 & 29,647,059 	\\
\hline
\end{tabular}
\end{table}

As can be observed from table \ref{table:Risk_module_SCR 2}, diversification effect accounts for approximately 17\% of the total (i.e. $\frac{6,218,424}{35,865,424}$). From table \ref{table:Risk_module_SCR 1}, it can be noticed that diversification effect is very high for the market and default risks. This depends on the correlation coefficient involved in the calculation.

\subsubsection{Non-Life Underwriting risk allocation} 

For the Non-Life Underwriting risk we can use the theoretical results provided in section \ref{sec:CORPO} to allocate Premium Risk, Reserve Risk and CAT Risk among LoB. Instead we use a market driven approach for the lapse risk referring to the best estimate of liabilities as risk driver, given its low importance in a non-life portfolio.

\paragraph{Allocation among sub-risks} \hspace{0pt} \\

From eq. \ref{EQ:subrisk_allocation_REF} we have that:

\begin{equation}
  SCR_{5j}^A= SCR_{5j} \cdot \frac{\sum\limits_{y=1}^{3}  SCR_{5y} \cdot \rho_{5x,5y}}{SCR_{5}} \cdot AR_{5}
\end{equation}

\begin{table}[!ht]
\centering
\caption{Non-Life Underwriting $SCR$ allocation among sub risks 1}
\label{table:Underwriting_SCR 1}
\begin{tabular}{l c r r r}
\hline
NL UdW sub-Risk & j & $SCR_{5j}$ & $SCR_{5j}^A$ &	$AR_{5j}$ \\ 	
\hline
Prem,Res risk    & 1 &  19,490,560    & 17,081,293    & 88\% \\
Lapse            & 2 &  552,645       & 12,137 	      & 2\%  \\
CAT              & 3 &  10,248,826    & 6,158,875     & 60\% \\
\hline
Total            &   &  30,292,030    & 23,252,305    & \\
\hline
\end{tabular}
\end{table} 
%
\begin{table}[!ht]
\centering
\caption{Non-Life Underwriting $SCR$ allocation among sub risks 2}
\label{table:Underwriting_SCR 2}
\begin{tabular}{l r}
\hline
$\sum_j SCR_{5j}$               & 30,292,030    \\
\hline
Sub-risk Diversification        & 6,218,424     \\
\hline
$SCR_5$                         & 24,188,911    \\
\hline
Risk-diversification            &  936,606      \\
\hline
$SCR_5^A$                       & 23.252.305    \\
\hline
\end{tabular}
\end{table} 

From Table \ref{table:Underwriting_SCR 1} it is worth noting that the cumulative benefit of a double diversification effect is involved. The first diversification derives from the risk module aggregation; the second one derives from the Non-Life sub-risk aggregation. Specifically based on the available data, the Lapse risk will suffer such a substantial reduction of the capital needed to cover it, as to become considered intangible.

\paragraph{Premium and Reserve Risk allocation among LoB} \hspace{0pt} \\

As previously carried out, the Premium \& Reserve Risk ($SCR_{51}^A=17,081,293$ and $SCR_{51}=19.490.560$) can be accurately allocated among each LoB using the general expression (\ref{EQ:r-level}):

\begin{equation} 
  SCR_{51k}^A= SCR_{51k} \cdot \frac{\sum\limits_{s=1}^{9}  SCR_{51s} \cdot \rho_{51k,51s}}{SCR_{51}} \cdot AR_{5} \cdot AR_{51} 
\end{equation}

\begin{table}[!ht]
\centering
\caption{Allocation of Premium  \& Reserve $SCR$ among LoBs 1}
\label{table:Prem_Res_SCR 1}
\begin{tabular}{l c l l l}
\hline
LoB (k) & $SCR_{51k}$   & $SCR_{51k}^A$ &	$AR_{51k}$ \\ 		
\hline
 1      &	 3,653,347 	&	 2,360,846 	&	65\%  \\
 2      &	 3,211,891 	&	 1,871,966 	&	58\%  \\
 3  	&	 2,779,696 	&	 1,497,000 	&	54\%  \\ 
 4 	    &	 2,102,026 	&	 997,678 	&	47\%  \\
 5 	    &	 3,586,055 	&	 2,113,211 	&	59\%  \\
 6 	    &	 1,061,883 	&	 521,882 	&	49\%  \\
 7 	    &	 2,642,109 	&	 1,596,281 	&	60\%  \\
 8 	    &	 1,609,509 	&	 854,498 	&	53\%  \\
 9 	    &	 6,830,006 	&	 5,267,930 	&	77\%  \\
\hline  
Total   &    27,476,524 &    17,081,293 &         \\
\hline
\end{tabular}
\end{table} 

\begin{table}[!ht]
\centering
\caption{Non-Life Underwriting $SCR$ allocation among sub risks 2}
\label{table:Prem_Res_SCR 2}
\begin{tabular}{l l}
\hline
$\sum_j SCR_{51j}$          & 27,476,524    \\
\hline
LoB Diversification         & 7,895,964     \\
\hline
$SCR_{51}$                  & 19,490,560    \\
\hline
subrisk-diversification     &   2,409,266   \\
\hline
$SCR_{51}^A$                & 17,081,293    \\
\hline
\end{tabular}
\end{table}

Furthermore, by means of \ref{EQ:r-level}, we can allocate, for each LoB, the total Premium \& Reserve risk dividing it into premium risk and reserve risk. We use the following formula:

\begin{equation}\begin{split}
  SCR_{51kz}^A &= SCR_{51kz} \cdot \frac{\sum\limits_{h=1}^{2}  SCR_{51kh} \cdot \rho_{51kh,51kz}}{SCR_{51k}} \cdot AR_{5} \\ & \cdot AR_{51} \cdot AR_{51k} 
\end{split}\end{equation}

where:

\begin{itemize}
    \item the third digit of the subscript identifies the LoB and is from 1 to 9;
    \item the fourth digit of the subscript identifies the premium risk (1) and reserve risk (2).
\end{itemize}

\begin{table}[!ht]
\centering
\caption{Allocation between Premium Risk and Reserve Risk}
\label{table:Prem_Res_All}
\begin{tabular}{l r r}
\hline
LoB (k) & $SCR_{51k1}^A$ & $SCR_{51k2}^A$   \\ 	
\hline
1	            &	 274,947 	&	 2,085,899   \\
2	            &	 447,103 	&	 1,424,863   \\
3	            &	 669,243 	&	 827,757     \\
4	            &	 218,669 	&	 779,009     \\
5	            &	 329,765 	&	 1,783,446   \\
6	            &	 221,695 	&	 300,188     \\
7	            &	 61,342 	&	 1,534,939   \\
8	            &	 669,081 	&	 185,418     \\
9	            &	 1,017,842 	&	 4,250,088   \\
\hline
SCR             &    3,909,685  &    13,171,608  \\
\hline

\end{tabular}
\end{table} 

\paragraph{CAT risk allocation} \hspace{0pt} \\

In the  standard  formula,  the  CAT  risk  is  defined  as  the  aggregation  of four sub risks: natural catastrophes, non-proportional reinsurance, man-made catastrophes, other catastrophes. Non-proportional reinsurance and other catastrophe risks have no further sub-risks. Natural catastrophes are divided in further sub-risks: windstorms, floods, earthquakes, hail, and subsidence perils. Man-made catastrophes include motor, marine, aviation, liability and credit. The capital requirement is obtained by means of a double level aggregation square root formula, so in order to obtain the capital allocation we use  (\ref{EQ:r-level}). In this case study we consider a portfolio with only natural catastrophe and man made catastrophes.
In following table we expose the results.

\begin{table}[!ht]
\centering
\caption{CAT Risk allocation 1}
\label{table:CAT Risk allocation 1}
\begin{tabular}{l c l l l}
\hline
CAT Risk & j & $SCR_{53j}$ & $SCR_{53j}^A$ &	$AR_{53j}$ \\ 	
\hline
Natural CAT         & 1     &   4.342.148       &  1.105.509    & 25\% \\
Man Made CAT        & 2     &   9.283.543       &  5.053.365    & 54\%  \\
\hline
Total               &       &   13.625.691      &  6.158.875    &     \\
\hline
\end{tabular}
\end{table} 

\begin{table}[!ht]
\centering
\caption{CAT Risk allocation 2}
\label{table:CAT Risk allocation 2}
\begin{tabular}{l r}
\hline
$\sum_j SCR_{53j}$              &   13.625.691   \\
\hline
Risk-level Diversification      &  3.376.866       \\
\hline
$SCR_{53}$                      &  10.248.826   \\
\hline
Upper level Diversification    &  4.089.951   \\
\hline
$SCR_{53}^A$                    &  6.158.875       \\
\hline
\end{tabular}
\end{table} 

\begin{table}[!ht]
\centering
\caption{Natural catastrophe risk allocation 1}
\label{table:CAT_NAT_All 1}
\begin{tabular}{l l l l l}
\hline
CAT Nat Risk & $SCR_{531y}$ & $SCR_{531y}^A$ &	$AR_{531y}$ \\ 	
\hline
Windstorm peril     &       -     	&        -    	 &    -           \\ 
Flood peril         &    2.272.544  &    260.360 	 &    13\%        \\
Earthquake peril    &    3.699.972 	&    802.694 	 &    22\%        \\
Hail peril          &        -     	&        -    	 &    -           \\
Subsidence peril    &        -     	&        -    	 &    -           \\
\hline
Total               &    5.972.516  &    1.105.509   &         \\
\hline
\end{tabular}
\end{table}

\begin{table}[!ht]
\centering
\caption{Natural catastrophe risk allocation 2}
\label{table:CAT_NAT_All 2}
\begin{tabular}{l r}
\hline
$\sum_j SCR_{531j}$             &  5.972.516     \\
\hline
Risk-level Diversification      &   1.630.368    \\
\hline
$SCR_{531}$                     &   4.342.148   \\
\hline
Upper level Diversification    &   3.236.639    \\
\hline
$SCR_{531}^A$                   &   1.105.509          \\
\hline
\end{tabular}
\end{table}

\begin{table}[!ht]
\centering
\caption{Man-Made catastrophe risk allocation 1}
\label{table:CAT_MANMADE_All 1}
\begin{tabular}{l l l l l}
\hline
Sub-risk & $SCR_{533y}$ & $SCR_{533y}^A$ &	$AR_{533y}$ \\ 	
\hline
Motor               &    2.391.787 	&    335.427 	 &    14\%      \\
Marine              &    3.438.637 	&    693.307 	 &    20\%      \\
Aviation            &        -     	&        -    	 &    -           \\
Fire                &    8.284.884 	&    4.024.631 	 &    49\%      \\ 
Liability           &        -     	&        -    	 &    -           \\
Credit              &        -     	&        -    	 &    -           \\
\hline
Total               &    14.115.308 &     5.053.365  &         \\
\hline
\end{tabular}
\end{table}

\begin{table}[!ht]
\centering
\caption{Man-Made catastrophe risk allocation 2}
\label{table:CAT_MANMADE_All 1}
\begin{tabular}{l r}
\hline
$\sum_j SCR_{532j}$             &    14.115.308     \\
\hline
Risk-level Diversification      &    4.831.765      \\
\hline
$SCR_{532}$                     &    9.283.543      \\
\hline
Upper level Diversification    &    4.230.178      \\
\hline
$SCR_{532}^A$                   &   5.053.365       \\
\hline
\end{tabular}
\end{table} 

Once the allocated SCR is obtained for each sub-risk for natural and man-made catastrophes, the allocation among LOBs is done considering a market driven approach and using the amount insured by LOBs as risk driver. In our case study, given the risks involved, we can consider each CAT sub-risk into its specific LOB: Motor (LOB 1), Marine (LOB 2), Other risks (LOB 4).

\subsection{Results}

We are now able to know the effective capital absorption of each Line of Business:

\begin{table}[!ht]
\centering
\caption{Results}
\label{table:Result1}
\begin{tabular}{l r r r r r}
\hline
LOB	&	Premium Risk	&	Reserve Risk	&	CAT	&	Lapse & Non-Life UdW Risk	 \\
\hline
 1   & 	 274.947 	 & 	 2.085.899 	 & 	 335.427 	    & 	 2.592 	 & 	 2.698.865 \\
 2 	 & 	 447.103 	 & 	 1.424.863 	 & 	 -   	        & 	 1.992 	 & 	 1.873.958 \\
 3 	 & 	 669.243 	 & 	 827.757 	 & 	 693.307 	    & 	 915 	 & 	 2.191.223 \\
 4 	 & 	 218.669 	 & 	 779.009 	 & 	 5.130.140 	    & 	 1.225 	 & 	 6.129.043 \\
 5 	 & 	 329.765 	 & 	 1.783.446 	 & 	 -   	        & 	 1.830 	 & 	 2.115.041 \\
 6 	 & 	 221.695 	 & 	 300.188 	 & 	 -   	        & 	 209 	 & 	 522.091 \\
 7 	 & 	 61.342 	 & 	 1.534.939 	 & 	 -   	        & 	 1.282 	 & 	 1.597.563 \\
 8 	 & 	 669.081 	 & 	 185.418 	 & 	 -   	        & 	 170 	 & 	 854.669 \\
 9 	 & 	 1.017.842 	 & 	 4.250.088 	 & 	 -   	        & 	 1.922 	 & 	 5.269.852 \\
\hline
Total 	&	 3.909.685   	 & 	 13.171.608 	 & 	 6.158.875 	 & 	 12.137 	 & 	 23.252.305

\end{tabular}
\end{table}

\section{Conclusion}

Based on the square root aggregation formula provided by EIOPA for Solvency 2 Capital Requirements, we have formalized a closed solution for the allocation problem coherent with the Euler Principle and applied in the standard formula framework. The outcomes avoid proxies in capital allocation up to the initial elements involved in capital aggregation where a square root aggregation formula, as the one represented in the paper, is applied. Furthermore, by means of specific proxies, we have shown how to perform an SCR allocation among LOBs. The availability of such information permits the shareholders to get in-depth knowledge about the capital absorption. In this sense, it enables to perform a series of strategic management actions that may be addressed for further research, for example:

\begin{itemize}
    \item the capital allocation optimization problem based on the return on absorbed capital, where performance is measured in terms of the Risk Adjusted Return on Equity (RORAC)
    \item the reduction of the Solvency Capital Requirement in order to respect risk appetite and risk tolerance  constraints.
\end{itemize}

%
%
%
%
%
%
%
%
%
%




\clearpage
\begin{thebibliography}{5}

\bibitem{Alb1}
  Albrecht P., 2004  
  \emph{Risk based capital allocation}. In: Encyclopedia of Actuarial Science. Wiley, Chichester.
\bibitem{Art1}
  Artzner P., Delbaen F., Eber J.M., Heath D., 1997  
  \emph{Thinking coherently}. RISK 10, 68-71.
\bibitem{Art2}
  Artzner P., Delbaen F., Eber J.M., Heath D., 1999
  \emph{Coherent measures of risk}. Mathematical Finance 9, 203-228.
\bibitem{BucDor}
  Buch A., Dorfleitner G., 2008
  \emph{Coherent risk measures, coherent capital allocations and the gradientallocation principle}. Insurance: Mathematics and Economics 42 235–242.
\bibitem{Dhaene1}
Dhaene J., Tsanakas A., Valdez E.A., Vanduffel S. \emph{The Journal of Risk and Insurance}, 2012, Vol. 79, No. 1, 1-28.
\bibitem{Den1}
  Denault M., 2001
  \emph{Coherent allocation of risk capital}. Journal of Risk 4, 7-21.
\bibitem{Denuit1}
	Denuit, M., Dhaene J., Goovaerts M., Kaas R., 2005. \emph{Actuarial Theory for Dependent Risks: Measures, Orders and Models.} JohnWiley and Sons, Chichester.

\bibitem{Denuit2}
 Denuit, M., Dhaene J., Goovaerts M., Kaas R., Laeven R., 2006. \emph{Risk measurement with the equivalent utility principles.} In Risk Measures: General Aspects and Applications, Editor
Ludger Rüschendorf, Statistics and Decisions,Vol. 24, No. 1, pgs. 1–26.

\bibitem{EIOPA-assumption}
	EIOPA, 25.07.2014, \emph{The underlying assumption in the standard formula for Solvency Capital Requirement calculation.} - https://eiopa.europa.eu.
\bibitem{EIOPA-Direttiva}
	EIOPA, 25.11.2009,\emph{Directive 2009/138/EC of the European parliament and of the council 25 November 2009 on the taking-up and pursuit of the business of Insurance and 		Reinsurance (Solvency II)}- https://eiopa.europa.eu.
\bibitem{EIOPA_TS}
	EIOPA, \emph{Technical Specification for the Preparatory Phase (Part I)} avaiable on https://eiopa.europa.eu/
\bibitem{EIOPA_Annex_K}
EIOPA ,\emph{see Annex k in "Annexes to the Technical Specification for Preparatory Phase".} - https://eiopa.europa.eu.


\bibitem{Filip1}
	Filipovic D., 2009. \emph{Multi-Level Risk Aggregation}. ASTIN Bulletin, 39, pgs. 565-575, doi:10.2143/AST.39.2.2044648.

\bibitem{McNeil1} 
	McNeil A., Frey R., Embrechts P., 2005. \emph{Quantitative RiskManagement.} Princeton University Press. Princeton series in Finance, Princeton,

\bibitem{Panjer2} 
	Panjer H., 2006 \emph{Operational Risk, Modeling Analytics}. John Wiley and Sons, Hoboken, NJ.
\bibitem{Sandstrom1} 
Sandstrom A., 2007, \emph{Solvency II: calibration for skewness}. Scandinavian Actuarial Journal.
\bibitem{Tas1}
	Tasche D., 1999, \emph{Risk contributions and performance measurement}. Working paper, Technische Universitat
Munchen, 1999.
\end{thebibliography}
\end{document}